\documentclass[preprint,10pt]{elsarticle}

\journal{Chaos, Solitons \& Fractals}

\usepackage{graphicx,bbm,setspace,amssymb,amsfonts,amsmath,mathtools,mathrsfs,stmaryrd,amsthm,bm,etoolbox,comment,hyperref,url,caption,subcaption,color,enumitem,algorithm,booktabs,microtype,nicefrac,lingmacros,nccmath,tree-dvips,tikz-cd}
\usepackage[noend]{algpseudocode}
\usepackage{algorithm}

\usepackage[colorinlistoftodos]{todonotes}
\usepackage[utf8]{inputenc} % allow utf-8 input
\usepackage[T1]{fontenc}    % use 8-bit T1 fonts

\DeclareMathOperator*{\argmax}{argmax}
\usepackage[noend]{algpseudocode}

\newtheorem{thm-defn}[theorem]{Theorem/Definition}

\theoremstyle{definition}

\theoremstyle{remark}

\DeclareMathOperator{\E}{\mathbb{E}}

\newcommand{\ignore}[1]{}{}

\begin{document}

\begin{frontmatter}

\title{Economic state classification and portfolio optimisation with application to stagflationary environments}
   
\author[label1]{Nick James} \ead{nick.james@unimelb.edu.au}
\author[label2]{Max Menzies} \ead{max.menzies@alumni.harvard.edu}
\author[label3]{Kevin Chin} % \ead{kevin.chin@arowanaco.com}
\address[label1]{School of Mathematics and Statistics, University of Melbourne, Victoria, Australia}
\address[label2]{Beijing Institute of Mathematical Sciences and Applications, Tsinghua University, Beijing, China}
\address[label3]{Arowana \& Co., NSW, 2060, Australia}

\begin{abstract}
Motivated by the current fears of a potentially stagflationary global economic environment, this paper uses new and recently introduced mathematical techniques to study multivariate time series pertaining to country inflation (CPI), economic growth (GDP) and equity index behaviours. We begin by assessing the temporal evolution among various economic phenomena, and complement this analysis with `economic driver analysis,' where we decouple country economic trajectories and determine what is most important in their association. Next, we study the temporal self-similarity of global inflation, growth and equity index returns to identify the most anomalous historic periods, and windows in the past that are most similar to current market dynamics. We then introduce a new algorithm to construct economic state classifications and compute an economic state integral, where countries are determined to belong in one of four candidate states based on their inflation and growth behaviours. Finally, we implement a decade-by-decade portfolio optimisation to determine which equity indices and portfolio assets have been most beneficial in maximising portfolio risk-adjusted returns in various market conditions. This could be of great interest to those looking for asset allocation guidance in the current period of high economic uncertainty.

\end{abstract}


\end{frontmatter}

\section{Introduction}
\label{Stagflation_background}

% Paragraph on stagflation
The term stagflation is highly definitional, but generally refers to an economic scenario where inflation is high and economic growth rate is stagnant (or slowing). Inflation itself has been relatively subdued for the past 20 years, providing an environment for business prosperity and growth in asset valuations among a range of asset classes, the most prominent of which is global equities. A current question of great relevance \cite{RayDalio_stagflation} is whether the currently elevated levels of inflation are transitory, or perhaps more structural (and in effect, pathological). This paper takes a data-driven and statistical approach to understand the drivers behind economic variables (growth, inflation and equity index returns), temporal self-similarity, the evolution of economic states/regimes, and the most effective assets for portfolio risk-adjusted returns on a decade-by-decade basis. 

% % Paragraph 2 Inflation/trajectories/offsets
Researchers have been interested in the relationship between economic growth and inflationary pressures for some time \cite{Siu2011,Mishkin1990,Fama1990,Mundell1963}. There is a substantial body of prior work that has focused on the application of principled econometric theory \cite{Tobin1965} to explore the evolutionary nature of inflationary behaviours. In the econophysics literature, however, there is limited work exploring the interplay between inflation, economic growth and equity index performance. In this paper, we use new and existing methodologies to better understand this relationship. This work could help economists and investment managers make better-informed decisions during the current economic climate, which is clouded by uncertainty surrounding the future of global inflation, growth and equity index dynamics.

% Financial market dynamics
Financial market dynamics, as a general topic of study, has been of great interest to researchers in applied mathematics, econophysics, econometrics and statistics. Temporal dynamics and correlation structures \citep{Fenn2011,Laloux1999,Mnnix2012} have garnered the interest of many researchers. Such studies are often accompanied by the study of evolutionary dynamics with techniques such as principal components analysis (PCA) \cite{Laloux1999,Kim2005,james_georg}, clustering \cite{Heckens2020,Jamesfincovid,arjun}, change point detection \cite{James2021_crypto,JamescryptoEq,james2021_MJW,James_crypto_physica_3,james2021_spectral}, entropy \cite{Wu2021,Chen2021}, chaotic systems \cite{Tacha2018,Cai2012,Szumiski2018}, topology \cite{Leibon2008} and various statistical modelling frameworks \cite{James2022_inflation,Ferreira2020}. Such methodologies have been applied to a wide variety of asset classes including equity markets \cite{Drod2001,Wilcox2007,Pan2007}, FX markets \cite{Ausloos2000,Gbarowski2019,Mikiewicz2021}, cryptocurrencies \cite{Stosic2019,Stosic2019_2,Manavi2020,Ferreira2020,Drod2018,Drod2019,Drod2020,Wtorek2020,james2021_crypto2,Chu2015,Lahmiri2018,Kondor2014,Bariviera2017,AlvarezRamirez2018} and debt-related instruments \cite{Driessen2003}. There is a litany of work where such techniques have been used in other domains including epidemiology \cite{James2021_geodesicWasserstein,james2021_CovidIndia,james2021_TVO}, extreme human behaviours \cite{ james2021_olympics} and energy \cite{james2021_hydrogen}. Readers interested in recent work related to temporal dynamics with various societal impacts on the economy should consult \cite{Sigaki2019,Perc_social_physics,Perc2019}.

% Optimisation
The topic of portfolio optimisation has been of great interest to quantitative finance researchers for an extended period of time \cite{Markowitz1952,Sharpe1966}. The core optimisation framework has been built upon in a variety of ways \cite{Almahdi2017,Calvo2014,Soleimani2009,Vercher2007,Bhansali2007,Moody2001,fister_two_2021,james2021_portfolio}. In this work, we implement the portfolio optimisation framework in a discrete, time-varying context where we may explore the effectiveness of various asset mixes in maximising investor portfolio Sharpe ratios. In particular, this paper explores the impact additional commodity-related asset classes (often used for hedging purposes) may have in diversifying a portfolio of equities allocated between various country equity indices.

This paper is structured as follows. In Section \ref{Economic_feature_similarity}, we investigate the temporal evolution in economic features' global variability and the drivers behind country economic features' association. In Section \ref{Temporal_self_similarity_analysis}, we explore the temporal self-similarity of each economic feature, and identify the most anomalous periods in history as well as periods in the past which may resemble current market dynamics. In Section \ref{economic_state_classification}, we introduce a new algorithm to determine the evolution of various country's economic profiles, and explore stagflationary tendencies with a novel economic state integral. We further investigate collective similarity by constructing a normalised inner product distance matrix and apply clustering to identify groupings among countries. In Section \ref{decade_portfolio_optimisation}, we apply a mean-variance portfolio optimiser to assess the effectiveness of various assets in maximising portfolio Sharpe ratio. We conclude in Section \ref{Conclusion}.

\section{Data}
\label{Data}

In this paper, we study quarterly and monthly time series data related to Consumer Price Index (CPI) inflation, gross domestic product (GDP) growth and equity index performance. We draw data from March 1960 to December 2021, a period of 60 years, for a collection of high profile economies: Australia, Canada, France, Germany, Italy, Japan, the United Kingdom and the United States. We select countries that could be classified as developed economies for the entirety of the period of analysis. For this reason, some notable emerging economies such as China, India and Brazil have been excluded from our analysis, as their economic conditions (especially dating back to the 1960s) may be incomparable with our chosen group of countries. In some experiments, monthly data is averaged over discrete quarters so quarterly time series may be compared fairly. In the final section of the paper, Section \ref{decade_portfolio_optimisation}, we study a collection of assets that includes the aforementioned country equity indices, gold spot price, oil spot price and the Commodity Research Bureau (CRB) commodity index (on a monthly basis).

\section{Economic feature similarity}
\label{Economic_feature_similarity}

Throughout this paper, the bulk of our analysis focuses on three multivariate economic time series pertaining to inflation (represented by CPI), economic growth (represented by GDP) and country equity indices (representing each country's equity market behaviours). We denote these time series $c_i(t)$, $g_i(t)$ and $e_i(t)$ respectively, where $i=1,...,N$ and $t=1,...,T$ refer to $N$ country indices and $T$ time points. 

\subsection{Evolutionary norm analysis}
\label{norm_analysis}

In this section, we explore the evolution in the total country variability for each economic feature. Given the stark differences in scale between countries' equity indices, before studying their collective evolution, we compute each countries' equity index log returns as follows:
\begin{align}
\tilde{e}_i(t) &= \log \left(\frac{e_i{(t)}}{e_i{(t-1)}}\right).
\end{align}
We then construct three univariate time series, where each one corresponds to the global state of inflation, growth and equity index performance, respectively. This construction is implemented as follows: 
\begin{align}
v^{c}(t) = \sum^N_{i=1} c_i(t), \\
v^{g}(t) = \sum^N_{i=1} g_i(t), \\
v^{\tilde{e}}(t) = \sum^N_{i=1} \tilde{e}_i(t),
\end{align}
where $v^c(t)$, $v^g(t)$ and $v^{\tilde{e}}(t)$ are time-varying functions. Importantly in our computation, we do not sum over the absolute value of each country's values as we wish to capture directional information (any feature could conceivably be positive or negative).

\begin{figure}
    \centering
    \begin{subfigure}[b]{0.49\textwidth}
        \includegraphics[width=\textwidth]{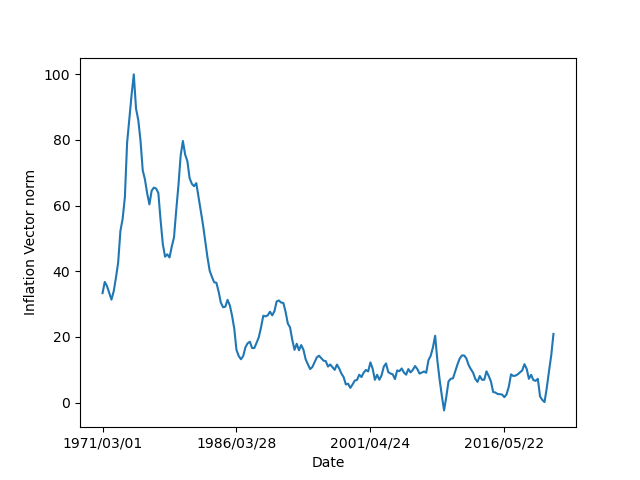}
        \caption{}
        \label{fig:Inflation_vector_norm}
    \end{subfigure}
    \begin{subfigure}[b]{0.49\textwidth}
        \includegraphics[width=\textwidth]{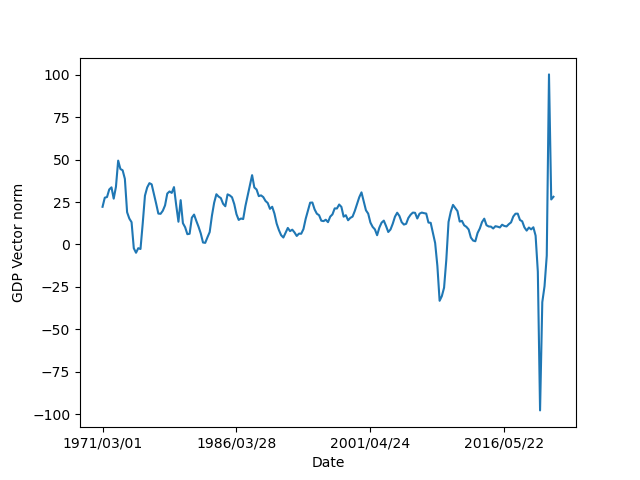}
        \caption{}
        \label{fig:GDP_vector_norm}
    \end{subfigure}
    \begin{subfigure}[b]{0.49\textwidth}
        \includegraphics[width=\textwidth]{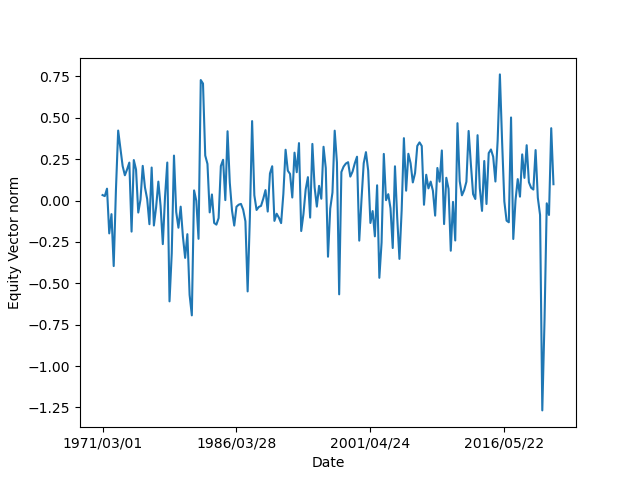}
        \caption{}
        \label{fig:Equity_vector_norm}
    \end{subfigure}
    \caption{Time-varying worldwide sums for CPI inflation, GDP growth and equity index returns, denoted by $v^c(t)$, $v^g(t)$ and $v^{\tilde{e}}(t)$ respectively. The figures demonstrate that in recent times, GDP growth and equity index returns have exhibited the most significant variability, largely due to the COVID-19 economic and financial market crises. The current acceleration in global inflation is alarming, and exhibits a similar trajectory to that seen in the 1970s.}
    \label{fig:Time_varying_norm}
\end{figure}

\begin{figure}
    \centering
    \begin{subfigure}[b]{0.49\textwidth}
        \includegraphics[width=\textwidth]{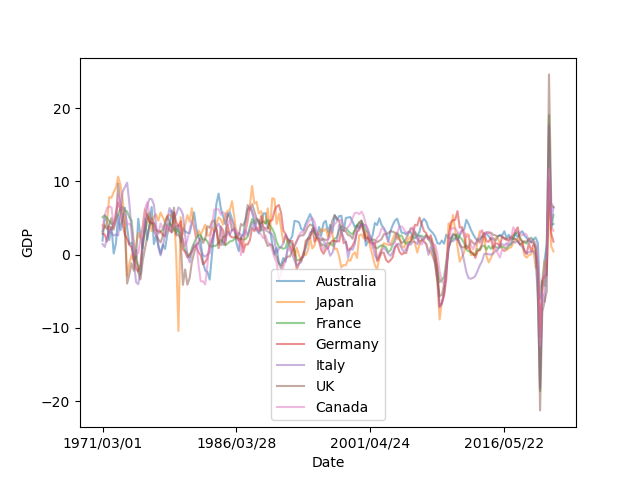}
        \caption{}
        \label{fig:GDP_trajectories}
    \end{subfigure}
    \begin{subfigure}[b]{0.49\textwidth}
        \includegraphics[width=\textwidth]{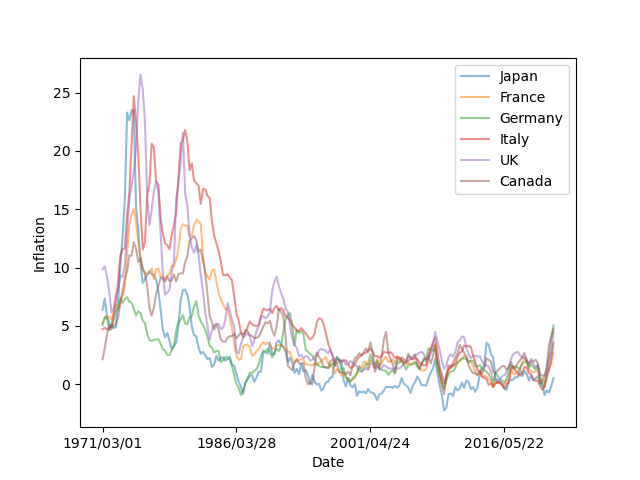}
        \caption{}
        \label{fig:Inflation_trajectories}
    \end{subfigure}
    \caption{Time-varying GDP and inflation levels by country. Australia has been excluded from the CPI inflation plot due to sparse data availability (quarterly rather than monthly data). Both figures indicate that over the past several years, GDP and inflation behaviours have been far more uniform than their historic variability. The one notable exception is Japan, which has produced anomalously low inflation relative to the other countries studied. }
    \label{fig:GDP_Inflation_trajectories}
\end{figure}

 Figure \ref{fig:Time_varying_norm} shows the time-varying collective behaviours among each economic feature studied over the past 60 years. Each time series displays noteworthy, but varied, behaviours of interest. We start by investigating the evolution of global inflation displayed in Figure \ref{fig:Inflation_vector_norm}. The figure tells an interesting story. After the significant levels of global inflation experienced during the 1970s and early 1980s (displayed for numerous countries in Figure \ref{fig:GDP_Inflation_trajectories}), there is a material reduction in global inflation from the early 2000s onwards. Although a small increase in global inflation occurred immediately prior to the global financial crisis (GFC), the prolonged crisis led to an immediate and prolonged retreat in country CPI levels. The past several years of data highlight a rapid acceleration in global CPI levels, a trajectory that has not been experienced since the early 1970s. Of course, the high levels of global inflation depicted are unsurprising, as they are widely discussed by Central Banks, investment and retail banks, fund managers and market commentators.

To complement our understanding of the temporal evolution in the economy, we turn to Figure \ref{fig:GDP_vector_norm}, which characterises the global effect of GDP growth. The two most notable attributes in the figure are the shocks corresponding to the GFC and the COVID-19 market crash (and its subsequent recovery). In particular, there is quite a clear distinction in the rate of the recovery following the GFC and the COVID-19 crash. The GFC market crash was significantly more prolonged and experienced a much slower recovery than that of COVID-19. The COVID-19 market crash exhibits an "down-up-down" pattern, highlighting the initial economic crash associated with the pandemic, the associated revival of GDP growth and the economy, and a final pullback as GDP levels begin to taper off.

Finally, we wish to study the relationship between the underlying economic variables and the effect they have on equity index behaviours. The time-varying total sum of equity index log returns is shown in Figure \ref{fig:Equity_vector_norm}. The most obvious feature in the figure is the sharp drawdown in returns corresponding to the COVID-19 market crash. This drop is especially pronounced, and the fundamental drivers behind such an intense and short-lived crash must be considered when we consider the current state of affairs. It is likely that in addition to restricted community mobility, business activity and the supply shocks experienced across many industries, a shift in investor composition toward more passive and factor-based investment products may have led to further indiscriminate selling during a time of crisis such as the pandemic.
 
 Viewing these figures in aggregate creates a holistic, albeit complex view of the economy and financial market's evolution. We have not seen a period in the last 60 years with such an unusual combination of features. These include: 
 \begin{enumerate}
     \item High levels of inflation, coming from a relatively low base in recent times.
     \item Economic growth (GDP) levels exhibiting such volatility post-pandemic.
     \item Equity markets displaying pronounced volatility on the back of (arguably) stretched valuation levels.
     \item An unprecedented level of investment manager money under management controlled by systematic, passive or factor-based investment strategies. It is likely that in times of economic and financial crisis, we will continue to see a widespread and somewhat mechanical selling of assets. 
 \end{enumerate}
 It is difficult to make an informed consideration of what may by likely moving forward, when the economic and financial circumstances surrounding a potential stagflationary environment are so unique.

\subsection{Economic "driver" analysis}
\label{driver_analysis}

In this section, we seek to answer the following question: \textit{is the country, or the specific economic feature, more important in driving similarity in behaviour?} To address this question, we consider each individual time series (inflation, growth and equity indices) for each country, to form a multivariate collection $f_j(t)$ where $j= 1,...,3N$ and again, $t=1,...,T$. 

Let $\|\mathbf{f}_{j}\|_1 = \sum^{T}_{t=1} |f_{j}(t)|$ be the $L^1$ norm of the respective economic or financial time series, and use this to normalise each trajectory. We denote this normalisation $\mathbf{T}^{f}_{j} = \frac{\mathbf{f}_j}{\|\mathbf{f}_j\|_1}$. The distance between two countries' trajectory vectors highlight the relative similarity in changes over time. We compute an \textit{economic driver distance matrix} as follows:

\begin{align}
\Omega_{jl}^{DR} = \|\mathbf{T}^f_i - \mathbf{T}^f_j\|_1.
\end{align}
\begin{figure}
    \centering
    \includegraphics[width=\textwidth]{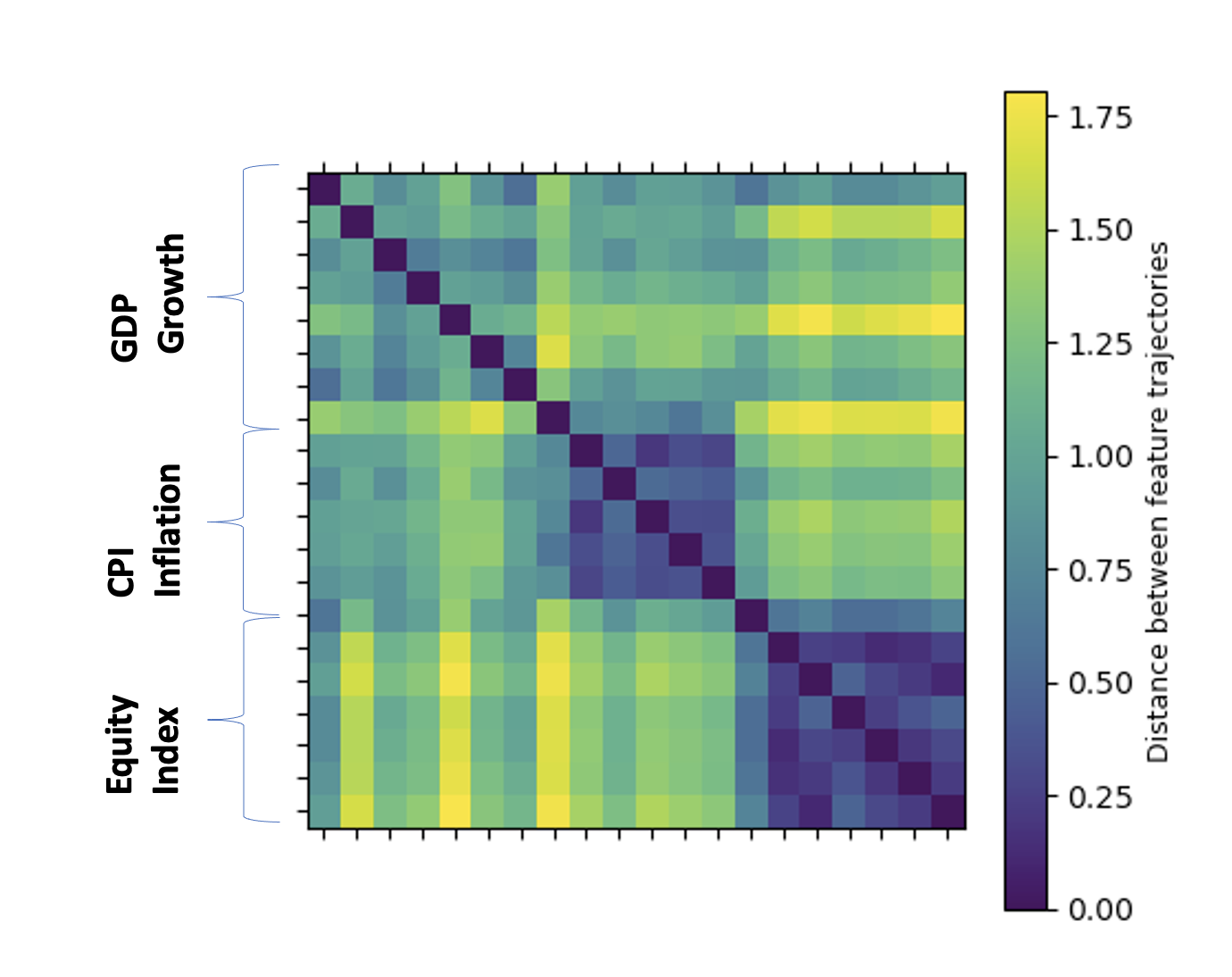}
    \caption{Economic driver distance matrix. The distance matrix shows three obvious groupings in evolutionary trajectories, driven by economic behaviours, rather than country features. That is, the inflation trajectories of two distinct countries are more likely to group together than the same countries' varying economic features (inflation, growth, and equity index behaviours). The more diffuse cluster structure of the two economic measures (GDP and inflation) demonstrates the strength of collective dynamics within equity markets. }
    \label{fig:Economic_driver_distance}
\end{figure}
The economic driver distance matrix, displayed in Figure \ref{fig:Economic_driver_distance}, reveals two interesting findings. First, it is clear that similarity between evolutionary trajectories is predominantly driven through economic behaviours, rather than specific countries' behaviours. That is, trajectories are more similar within one class of behaviours (inflation, growth, equity indices) than within one country's time series. This highlights the systemic nature of such economic phenomena, and may be driven through the strength of the collective dynamics in the global economy. The second finding of note, is the varying degree of self-similarity among different economic features. Equity index returns clearly display the strongest degree of self-similarity (visible in the low distances between equity returns), followed by inflation trajectories and then GDP growth trajectories. This finding may suggest that, at least among the economies studied in this paper, inflation behaviours are more globally systemic than economic growth patterns.

\section{Temporal self-similarity analysis}
\label{Temporal_self_similarity_analysis}

In this section, we explore how similar each economic feature is to periods in the past. We refer to this phenomenon as \textit{temporal self-similarity}. The two primary motivating questions we seek to answer are as follows. First, which features exhibit, or have exhibited, the most extreme behaviours relative to current market conditions. Second, for each feature, which period in the past most closely resembles the current time period. We explore each economic feature (GDP growth, CPI inflation and country equity index returns) separately. First, we wish to study the similarity in vector norms between all possible points in time. For the sake of exposition, we detail how this would be computed on the inflation time series. Let $\mathbf{c}(t)=(c_1(t),...,c_N(t)) \in \mathbb{R}^N$ be the vector of all inflation data at a time $t$. We compute an $L^1$ distance (again denoted $\| \cdot \|_1$) between such vectors of country inflation data between every time $t=1,...,T$. This produces a $T \times T$ distance matrix 
\begin{align}
d^{c}(s,t) = \| \mathbf{c}(s) - \mathbf{c}(t)\|_1  \forall s,t \in \{1,...,T\}.    
\end{align}
 We compute analogous distance matrices for growth and equity index performance, which we denote $d^{g}(s,t)$ and $d^{ \tilde{e}}(s,t)$.

\begin{figure}
    \centering
    \includegraphics[width=\textwidth]{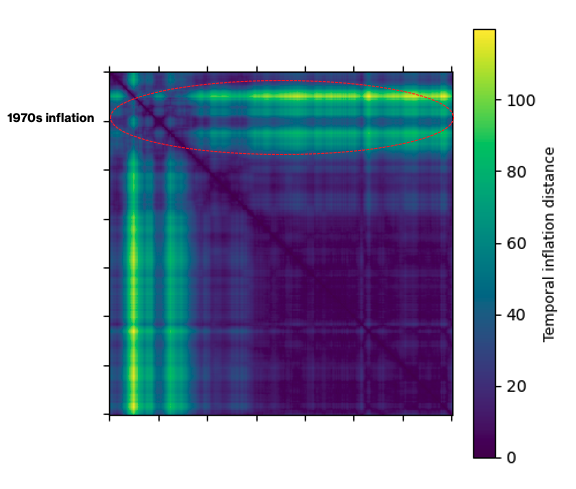}
    \caption{$d^{c}(s,t)$ distance matrix computing $L^1$ distance between vectors of country CPI values at all possible points in time. The most anomalous period in time corresponds to the high global inflation levels in the 1970s. Beyond this point in time, there is significantly less temporal distance between global inflation behaviours.}
    \label{fig:D_st_inflation}
\end{figure}

\begin{figure}
    \centering
    \includegraphics[width=\textwidth]{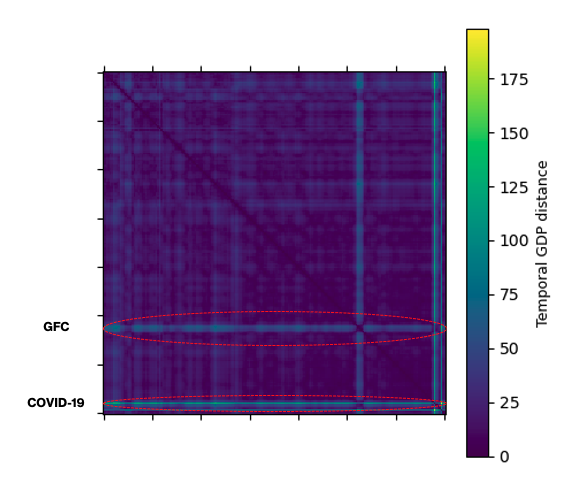}
    \caption{$d^{g}(s,t)$ distance matrix computing $L^1$ distance between vectors of country GDP values at all possible points in time. The two most anomalous periods correspond to the GFC and the COVID-19 market crash. It is interesting to note that the dot-com bubble, which is widely considered to be a major economic and financial market crash, is not easily identified when measuring temporal GDP distance.}
    \label{fig:D_st_gdp}
\end{figure}

\begin{figure}
    \centering
    \includegraphics[width=\textwidth]{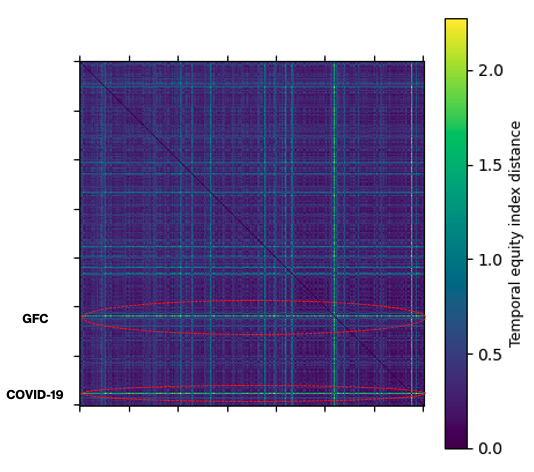}
    \caption{$d^{\tilde{e}}(s,t)$ distance matrix computing $L^1$ distance between vectors of country equity index returns at all possible points in time. The two most anomalous periods correspond to the GFC and COVID-19 market crash. Unlike our temporal GDP analysis, the dot-com bubble is identified in this study, reflecting a more significant equity market crash, rather than a systemic fault in the underlying economy. Both the GFC and COVID-19 market crises exhibit greater temporal equity index distance than the dot-com bubble, relative to market steady state.}
    \label{fig:D_st_equity}
\end{figure}

Each distance matrix reveals the temporal self-similarity displayed by each economic/financial feature analysed. 
% Inflation
We start by studying Figure \ref{fig:D_st_inflation}, which shows the $L^1$ distance between vectors of country CPI values at all possible points in time. The figure reveals one notably anomalous period, corresponding to the 1970s. This is consistent with our earlier discussion in Section \ref{Economic_feature_similarity}, as the 1970s was a period of especially high global inflation. The part of the distance matrix corresponding to this period in time is indicated by a red, dashed elliptical annotation. The second aspect of the distance matrix of considerable interest is the period of the late 1990s and early 2000s corresponding to low levels of global inflation. This is also consistent with Section \ref{Economic_feature_similarity}, which highlights a subdued period of inflation from the early 2000s onward (following a slow transition downward from the high levels of inflation during the 1970s).

% Growth
We then turn to Figure \ref{fig:D_st_gdp}, which displays the $L^1$ distance between vectors of country GDP values at all possible points in time. There are two notably different periods of time, corresponding to the GFC and COVID-19 market crash. Again, these periods in time are marked by a red, dashed elliptical annotation. This finding is in agreement with Section \ref{Economic_feature_similarity}. Outside these two periods, the temporal self-similarity in GDP growth is remarkably constant - with the distribution of global GDP levels exhibiting limited deviation over time. 

% Equity d_st
Finally, we explore Figure \ref{fig:D_st_equity}, which shows the temporal self-similarity between vectors of country equity returns. Again, the two most prominently anomalous periods in time are the GFC and COVID-19 market crash. Both periods are marked by red, dashed elliptical annotations. This finding may be indicative of the increase in financial market correlations during crises over the past several decades. In particular, periods of economic distress are accompanied by a dramatic increase in the average correlation coefficient among equity market returns. There is a litany of work that discusses such phenomena.

When viewed together, these three figures highlight an interesting story. Both global GDP and country equity indices are characterised by relatively homogenous behaviours outside the two most recent and severe market crises (GFC and COVID-19). By contrast, these periods did not impact global inflation levels, with significant homogeneity exhibited beyond the early 2000s. However, the unprecedented high levels of inflation experienced during the 1970s are shown to be materially different to all other periods in time under consideration.

\section{Economic state classification}
\label{economic_state_classification}

In this section, we develop a new algorithm to dynamically determine the economic state of a country at any candidate point in time. Given our particular emphasis on studying the time-varying association between inflation and growth on a country-by-country basis, we restrict our study to each country's inflation (CPI) and growth (GDP) time series. We begin with the (strong) assumption, that a country may be in one of 4 potential states:
\begin{enumerate}
    \item Ascending growth: Above threshold growth and positive inflation
    \item Descending growth: Above threshold growth and negative inflation (deflation)
    \item Stagflation: Below threshold growth and declining inflation
    \item Contraction: Below threshold growth and deflation
\end{enumerate}
Obviously, these states are highly dependent on the definition of "threshold growth". In our proceeding experiments, for any country $i$, we define threshold growth to be $\mu_i^g - \sigma_i^g$ where $\mu_i^g$ and $\sigma_i^g$ are the mean GDP level and standard deviation GDP level for each country. Intuitively, we are saying that above threshold growth occurs approximately five sixths of the time, and below threshold occurs the exceptional one sixth of the time when growth is more than one standard deviation from the mean (to the left). Algorithm \ref{algorithm} outlines the economic state classification, which is conditional on a country's level of GDP (relative to its average) and CPI inflation.

\begin{figure}
    \centering
    \begin{subfigure}[b]{0.49\textwidth}
        \includegraphics[width=\textwidth]{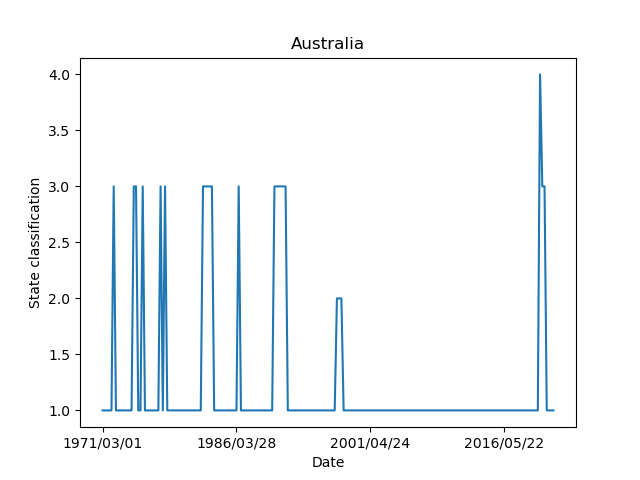}
        \caption{}
        \label{fig:Stagflation_Australia}
    \end{subfigure}
    \begin{subfigure}[b]{0.49\textwidth}
        \includegraphics[width=\textwidth]{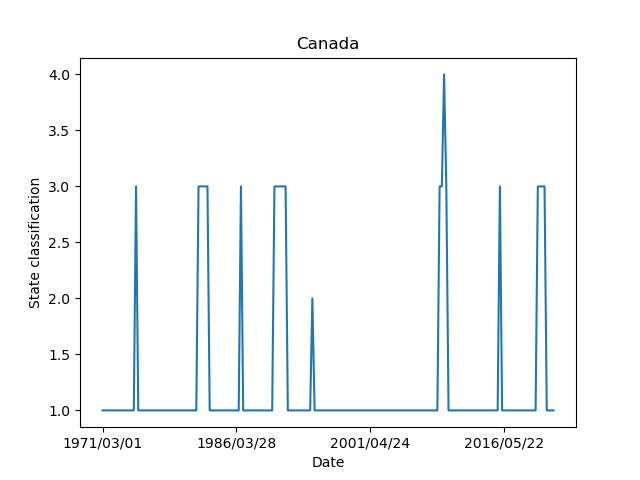}
        \caption{}
        \label{fig:Stagflation_Canada}
    \end{subfigure}
    \begin{subfigure}[b]{0.49\textwidth}
        \includegraphics[width=\textwidth]{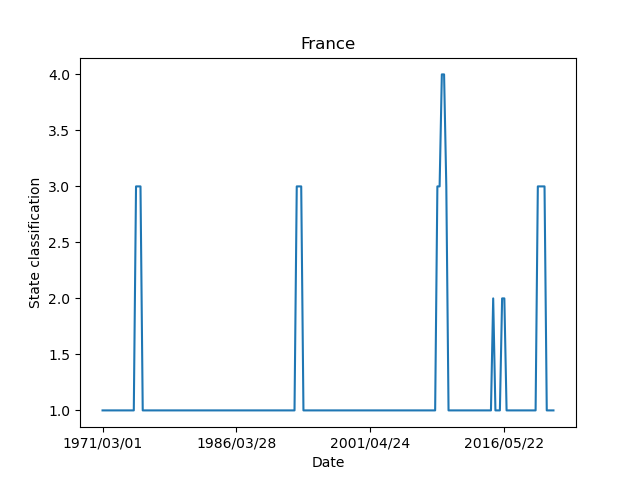}
        \caption{}
        \label{fig:Stagflation_France}
    \end{subfigure}
    \begin{subfigure}[b]{0.49\textwidth}
        \includegraphics[width=\textwidth]{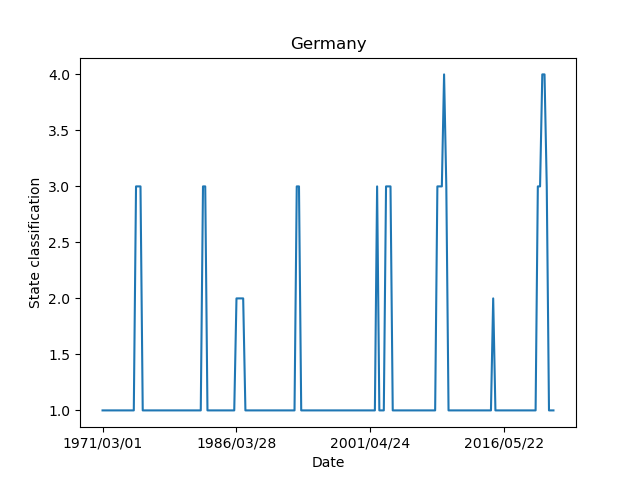}
        \caption{}
        \label{fig:Stagflation_Germany}
    \end{subfigure}
    \begin{subfigure}[b]{0.49\textwidth}
        \includegraphics[width=\textwidth]{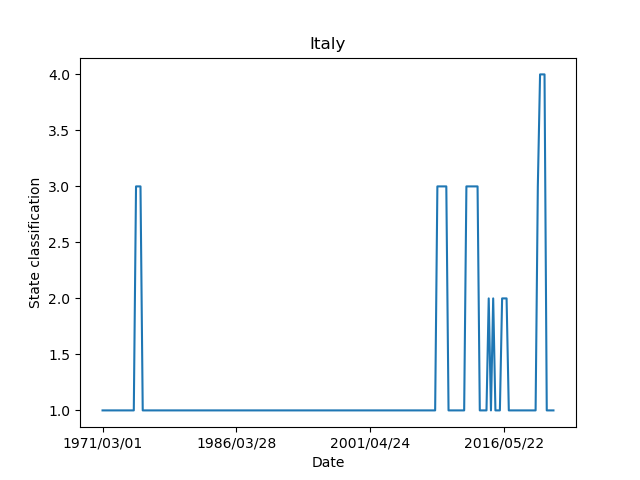}
        \caption{}
        \label{fig:Stagflation_Italy}
    \end{subfigure}
    \begin{subfigure}[b]{0.49\textwidth}
        \includegraphics[width=\textwidth]{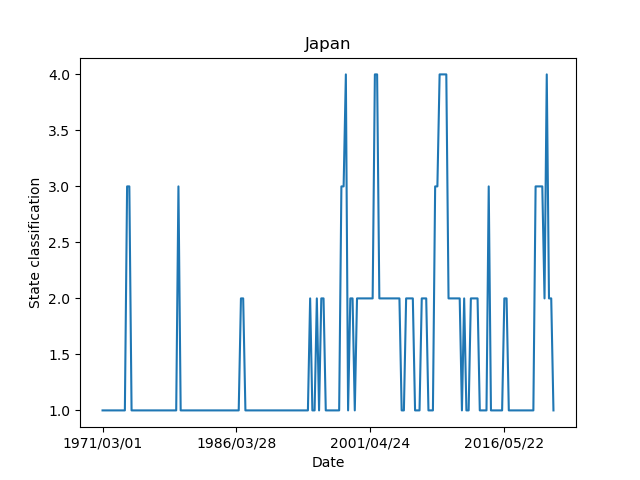}
        \caption{}
        \label{fig:Stagflation_Japan}
    \end{subfigure}
        \begin{subfigure}[b]{0.49\textwidth}
        \includegraphics[width=\textwidth]{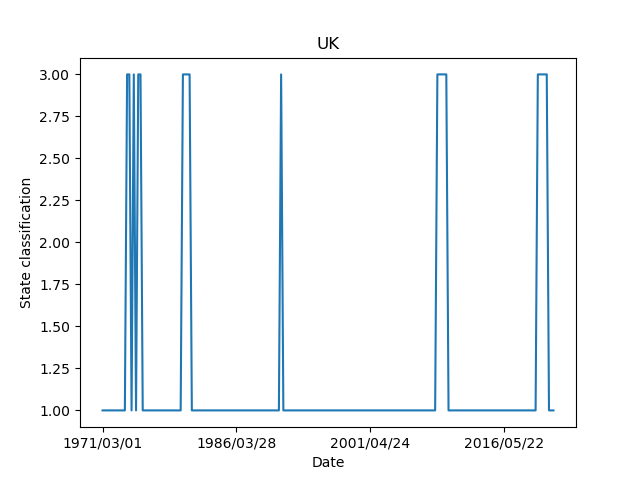}
        \caption{}
        \label{fig:Stagflation_UK}
    \end{subfigure}
    \begin{subfigure}[b]{0.49\textwidth}
        \includegraphics[width=\textwidth]{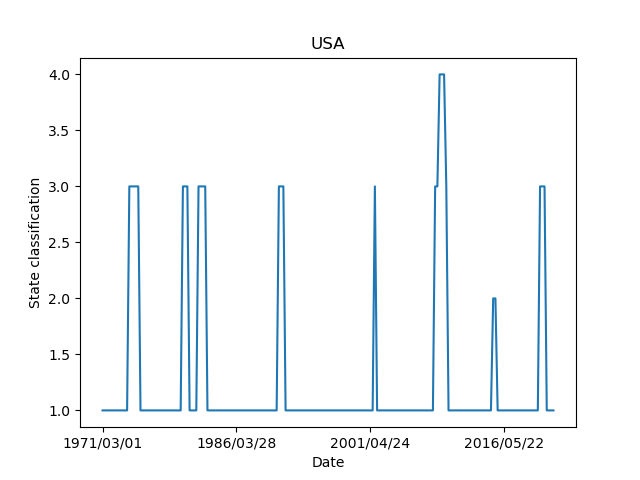}
        \caption{}
        \label{fig:Stagflation_USA}
    \end{subfigure}
    \caption{Country economic state classification time series $\mathcal{S}_i(t)$ computed by Algorithm 1 for (a) Australia, (b) Canada, (c) France, (d) Germany, (e) Italy, (f) Japan, (g) the UK and (h) the US. Japan propagates toward state 4 most frequently, which is consistent with its high $\Delta^{(i)}$ score. There is evidently high variability among country economic state behaviours, with limited concurrence of worse conditions.}
    \label{fig:Country_economic_state_time_series}
\end{figure}

\begin{algorithm}[H]%[tb]
	\caption{Economic state classification and economic integral computation} % Economic state classification algorithm
	\label{algorithm}
	\begin{algorithmic}[1]	
	    \State Initialise empty list for country state integral, $\Delta^{i}$
	    \For{$i=1 \,\text{ to }\, N$}
	    \State Initialise empty list for economic state time series, $\mathcal{S}_i(t)$
	    \State Compute mean of country GDP time series, denoted $\mu_i^g$
	    \State Compute standard deviation of country GDP time series, denoted $\sigma_i^g$
	    \For{$t=1 \,\text{ to }\, T$}
	    \If {$g_i(t) > (\mu_i^g - \sigma^g_i$) and $c_i(t) > 0$}
	    \State State declared as \textit{ascending growth}
	    \State $\mathcal{S}_i(t) = 1$. Append to list.
	    \EndIf
	    \If {$g_i(t) > (\mu_i^g - \sigma^g_i$) and $c_i(t) \leq 0$}
	    \State State declared as \textit{descending growth}
	    \State $\mathcal{S}_i(t) = 2$. Append to list.
	    \EndIf
	    \If {$g_i(t) \leq (\mu_i^g - \sigma^g_i$) and $c_i(t) > 0$}
	    \State State declared as \textit{stagflation}
	    \State $\mathcal{S}_i(t) = 3$. Append to list.
	    \EndIf
	    \If {$g_i(t) \leq (\mu_i^g - \sigma^g_i$) and $c_i(t) \leq 0$}
	    \State State declared as \textit{contraction}
	    \State $\mathcal{S}_i(t) = 4$. Append to list.
	    \EndIf
	    \EndFor 
	    \State Compute $\Delta^{(i)} =  \frac{1}{T}  \sum^T_{t=1} \mathcal{S}_i(t)$.
	    \State Output each country's economic state classification time series, $\mathcal{S}_i(t)$ and \emph{economic state integral} $\Delta^{(i)}$.
	    \EndFor 
	\end{algorithmic}
\end{algorithm}

For each country $i$, we compute $\mathcal{S}_i(t)$ and $\Delta^i$, as defined in the algorithm. The former gives a time-varying economic state classification, and is displayed for various countries in Figure \ref{fig:Country_economic_state_time_series}. We remark that higher values of $S_i(t)$ (according to the given ordering of states) indicate less prosperous economic conditions and greater economic instability. Thus, $\Delta_i$, termed the  \emph{state economic integral}, provides an overall measure of a country's tendency to enter stagflationary, deflationary or recessionary periods. Such values are recorded in Table \ref{tab:Economic_state_integral}.

\begin{table}
\centering
\begin{tabular}{ |p{2.9cm}||p{2cm}|}
 \hline
 \multicolumn{2}{|c|}{Economic state integral} \\
 \hline
 Country & $\Delta^{(i)}$ \\
 \hline
 Australia & 1.22  \\
 Japan & 1.47  \\
 France & 1.17 \\
 Germany & 1.24 \\
 Italy & 1.21 \\
 UK & 1.19  \\
 Canada & 1.22 \\
 USA & 1.27  \\
\hline
\end{tabular}
\caption{Economic state integral, $\Delta^{(i)}$ computed from each countries' economic state classification time series $\mathcal{S}_i(t)$. The scores indicate that Japan has been the economy with the strongest stagflationary or recessionary tendencies. This is consistent with the history of economic and financial market development over the past 60 years.}
\label{tab:Economic_state_integral}
\end{table}

Table \ref{tab:Economic_state_integral} has two primary takeaways. The first of which is the relative degree of similarity among most countries' $\Delta^{(i)}$ scores. This indicates the systemic nature of both CPI inflation and GDP growth, reflecting that most countries share a similar tendency for deflationary (et al.) market conditions. The most notable outlier is Japan. Japan's economic state integral is materially different to the rest of the country collection, and may reflect the economic stagnation experienced by the country during the 1990s.

\begin{figure}
    \centering
    \includegraphics[width=\textwidth]{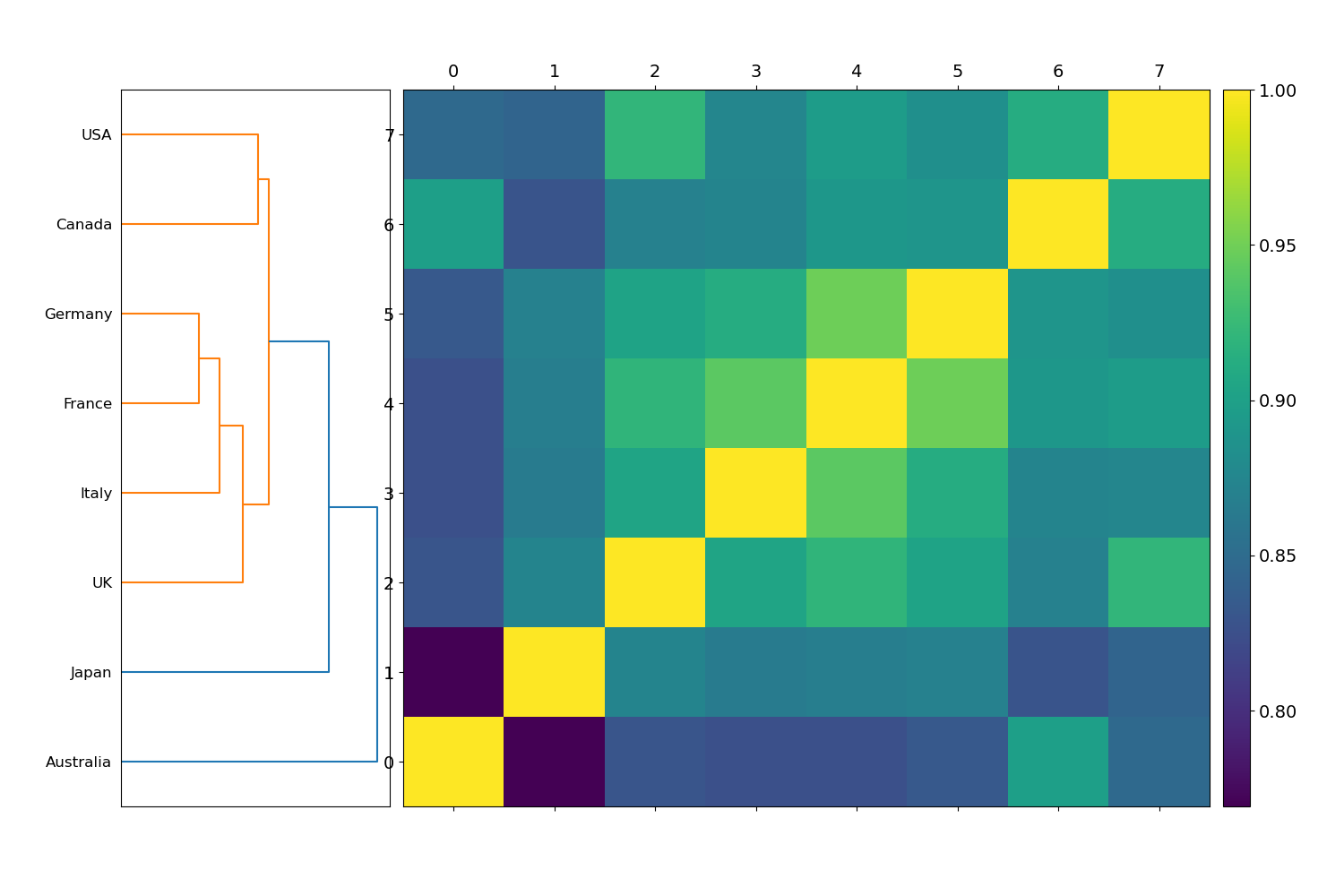}
    \caption{Hierarchical clustering applied to our economic state matrix $\Omega^{\mathcal{S}}$. The algorithm determines the existence of $k^{\mathcal{S}} = 2 $ clusters, with Japan and Australia displaying materially different behaviours to the remaining countries. Indeed, as seen in Figure \ref{fig:Country_economic_state_time_series}, Australia exhibits an unusually long period of state 1 since 2000, while Japan experiences the opposite, the most frequent time spent in states 2-4. This is most likely in relation to these countries' inflationary behaviours, which have been studied in previous work.}
    \label{fig:economic_state_distance_matrix}
\end{figure}

To support this computation, we compute a normalised inner product matrix between each of our country's state classification time series as follows: 
\begin{align}
\Omega^{\mathcal{S}}_{ij} = \frac{<\mathcal{S}^{(i)}, \mathcal{S}^{(j)}>}{||\mathcal{S}^{(i)}||_2 ||\mathcal{S}^{(j)}||_2}.    
\end{align}
This yields an $N \times N$ distance matrix that captures the similarity in the interplay between GDP growth and inflation among our countries under investigation. We term this matrix an \textit{economic state matrix}. To group countries with similar evolutionary behaviours, we apply hierarchical clustering to our matrix $\Omega^{\mathcal{S}}$, and analyse the estimated number of clusters $k^{\mathcal{S}}$, and the association between various countries. The resulting dendrogram is shown in Figure \ref{fig:economic_state_distance_matrix}. There we observe $k^{\mathcal{S}}=2$ clusters, with Australia and Japan forming an outlier cluster of their own. Japan's outlier status in its economic conditions has already been noted; Australia's outlier status is due to an exceptionally prolonged period of stable market conditions since the year 2000, as seen by the long constant period of economic state 1 in Figure \ref{fig:Stagflation_Australia}.

\subsection{Markov transition matrices}

We build upon the state classification time series of each country $\mathcal{S}_i(t)$ by studying the transition probability matrix of each country. For each country $i$, we generate a transition probability matrix $P^{(i)}$ between our $r=4$ states given by
\begin{align}
    P^{(i)} = \left[\begin{array}{cccc}
p^{(i)}_{11} & p^{(i)}_{12} & p^{(i)}_{13} & p^{(i)}_{14}	\\
p^{(i)}_{21} & p^{(i)}_{22} & p^{(i)}_{23} & p^{(i)}_{24}	\\
p^{(i)}_{31} & p^{(i)}_{32} & p^{(i)}_{33} & p^{(i)}_{34}	\\
p^{(i)}_{41} & p^{(i)}_{42} & p^{(i)}_{43} & p^{(i)}_{44}
\end{array}\right].
\end{align}
Each individual $j$th row consists of the empirical frequencies $p_{jk}$ that the state $j$ will be followed by one of states $k=1,2,3,4.$ Thus, the matrix $P$ is stochastic, with rows consistenting of non-negative entries that sum to 1. To highlight the difference in economic state behaviours, we contrast the transition probability matrices of Australia and the United States (US). 
\begin{align}
    P^{AUS} = \left[\begin{array}{cccc}
0.9438 & 0.0056 & 0.045 & 0.0056 \\
0.33 & 0.67 & 0.00 & 0.00 \\
0.45 & 0.00 & 0.55 & 0.00 \\
0.00 & 0.00 & 1.00 & 0.00 
\end{array}\right].
\end{align}

\begin{align}
    P^{US} = \left[\begin{array}{cccc}
0.9543 & 0.0057 & 0.04 & 0.00 \\
0.50 & 0.50 & 0.00 & 0.00 \\
0.318 & 0.00 & 0.6363 & 0.045 \\
0.00 & 0.00 & 0.33 & 0.67 
\end{array}\right].
\end{align}

Both countries display broad similarity in their tendency to stay in state 1 of ascending growth, with Australia exhibiting $P(S_{t+1} = 1 | S_t = 1) = 0.9438$ which is comparable to the US $P(S_{t+1} = 1 | S_t = 1) = 0.9543$. The second major difference is the distinction in Australia and the US' apparent `stickiness' in state 4 - periods of contraction. The contrasting transition probability matrices suggest that the US tends to have more prolonged recessions, with $P(S_{t+1}=4|S_t=4) = 0.67$ and $P(S_{t+1}=3|S_t=4) = 0.33$. By contrast, Australia exhibits a transition probability $P(S_{t+1}=3|S_t=4) = 1$, implying that whenever Australia has experienced a period of recession, it is quick to transition into a better economic state. These transition probabilities are consistent with the history of these two countries' economies over the last several years, in particular. For example, the GFC had a much more significant impact on the US economy than that of Australia, and can be seen in Figure \ref{fig:Stagflation_USA}. During the global recession, the Australian economy was largely protected due to pre-existing strong governance across the big primary financial institutions \cite{ausbigfourbanks} and a boom in the resources sector where the Australian economy is heavily leveraged \cite{ausresourceboom}.

\section{Decade-by-decade portfolio optimisation}
\label{decade_portfolio_optimisation}

We conclude our analysis with a dynamic portfolio optimisation exercise whereby we wish to determine the most essential assets in maximising an investor's Sharpe ratio in varying market conditions. Let our collection of asset prices be $a_k(t)$ where $k=1,...,K$ indexes our assets under study. This collection includes the country indices of Japan, France, Germany, Italy, United Kingdom, Canada and the United States, gold, oil and the CRB Commodity index. We sequentially apply our optimisation algorithm for 10-year periods between 1960-1970, 1970-1980,...,2010-2020. One must note that due to limitations in our dataset, oil is omitted from the first two decades' optimisation experiments. 
We begin by computing the log returns for each asset as follows:
\begin{align}
    \tilde{a}_k(t) = \log \bigg(    \frac{a_k(t)}{a_k(t-1)}     \bigg)
\end{align}

As mentioned above, we wish to conduct our experiment on a decade-by-decade basis, so we choose a window of $\tau=120$ (months) to evaluate our optimisation results. Intuitively, this length is chosen so that we examine the trailing 10 years of data, and study the evolution in optimal portfolio weights as we propagate forward in time. The optimisation objective function is constructed as follows:
\begin{align}
\argmax_{w_1,...,w_K} \frac{\E (R_{p} {[t-\tau : t])}}{\sigma_{p}^{2}[t-\tau:t]},  \forall t \in \{120, 240, ..., 720\}, \text{ where}\\
    \E (R_{p}) = \sum^{K}_{k=1} w_{k} \tilde{a}_{k}, \text{ and} \\
    \sigma^2_{p} = \boldsymbol{w}^{T} \Sigma \boldsymbol{w}.
\end{align}
The notation $[t-\tau:t]$ signifies that the Sharpe ratio is computed and optimised sequentially over time intervals $[t-\tau,t]$. We restrict each asset's portfolio weight $0.025 \leq w_{k} \leq 0.4, k = 1,...,K$ and assume that the portfolio is always fully invested in a long-only capacity $\sum^{K}_{k=1} w_{k} = 1$. We are judicious in our choice of restricting asset values between 2.5\% and 40\% of the total portfolio size. Given that we are aiming to simulate portfolio allocation decisions of an asset allocation firm (such as an endowment, pension fund, and so on), we place bounds on the weights of our candidate investments that may be comparable to a real-world investment position sizing \cite{Russellpolicy}. Within the Markowitz portfolio optimisation framework, bounds are generally needed on asset weights to alleviate the risk of generating trivial solutions. Similarly, should we enforce bounds that are excessively tight, we may fail to optimise over a sufficiently large space such that we can reach an appropriate optimal solution.

At each point in time, the optimisation generates a vector of optimal portfolio weights that we denote $w_k^{*}$. We apply these constraints to simulate the dynamics of a global asset allocator, who may seek to diversify their investments geographically and have a long-term holding period with respect to candidate investments. The results in Table \ref{tab:Optimal_portfolio_weights_} show the evolution in optimal portfolio weights when seeking to maximise the portfolio Sharpe ratio.

\begin{table}
\centering
\begin{tabular}{ |p{0.8cm}||p{0.65cm}|p{0.65cm}|p{0.65cm}|p{0.65cm}|p{0.65cm}|p{0.65cm}|p{0.65cm}|p{0.65cm}|p{0.65cm}|p{0.65cm}|}
 \hline
 \multicolumn{11}{|c|}{Dynamic optimal portfolio weights} \\
 \hline
 Time & Gold & Oil & CRB & JPN & FR & GER & ITY & UK & CAN & US \\
 \hline
 1960s & 0.025 & N/A & 0.025 & 0.33 & 0.025 & 0.025 & 0.025 & 0.18 & 0.34 & 0.025\\
 1970s & 0.4 & N/A & 0.22 & 0.18 & 0.025 & 0.025 & 0.025 & 0.07 & 0.025 & 0.025 \\
 1980s & 0.025 & 0.025 & 0.025 & 0.4 & 0.025 & 0.03 & 0.17 & 0.25 & 0.025 & 0.025 \\
 1990s & 0.025 & 0.082 & 0.025 & 0.025 & 0.025 & 0.025 & 0.025 & 0.34 & 0.025 & 0.40 \\
 2000s & 0.40 & 0.025 & 0.39 & 0.025 & 0.025 & 0.025 & 0.025 & 0.025 & 0.025 & 0.036 \\
 2010s & 0.35 & 0.025 & 0.025 & 0.062 & 0.025 & 0.04 & 0.025 & 0.025 & 0.025 & 0.40 \\
\hline
\end{tabular}
\caption{Decade-by-decade optimal portfolio weights. Abbreviations in the table heading are as follows: CRB: CRB Commodity index, JPN - Japan equity index, FR - French equity index, GER - German equity index, ITY - Italy equity index, UK - UK equity index, CAN - Canada equity index, US - US equity index.}
\label{tab:Optimal_portfolio_weights_}
\end{table}

Table \ref{tab:Optimal_portfolio_weights_} tells an interesting story concerning the most important countries and commodities to allocate capital towards in various market conditions. During the 1960-1970 period, the two equity indices that should have received the largest capital allocation were Japan and Canada, with 33\% and 34\% of total money invested, respectively. During this period, both equity indices experienced strong growth. During the 1970-1980 period, however, optimal weight allocations were markedly different. This decade was characterised by high inflation, and unsurprisingly the two assets that receive the largest optimal capital allocation are gold and the CRB Commodity index, comprising 40\% (the upper bound of the weight constraint) and 22\% respectively. During the 1980-1990 and 1990-2000 decades the most effective assets in maximising portfolio Sharpe ratio were the Japan and US equity indices, both comprising 40\% of total allocated capital. During the 2000-2010 period, more than 79\% of total capital should be allocated toward gold (40\%) and commodities (39\%) respectively. Finally, during the 2010-2020 period, the two most critical assets in maximising risk-adjusted returns were the US equity index (40\%) and gold (35\%), respectively. 

We proceed by computing the average optimal portfolio weight for each asset. This calculation reveals that gold has the highest average optimal portfolio weight $\sim$ 20.4\%, which may suggest that it remains an essential inclusion in equity portfolios during varying market dynamics. Although the Japan equity index has an average optimal weight allocation of $\sim$ 17\%, the optimal allocation has declined significantly in recent decades due to the significant economic pressures the country has faced, and continues to face.

\section{Conclusion}
\label{Conclusion}

This paper uses various data, mathematical techniques and frameworks to provide a holistic view of inflation, and its impact on equity markets and (predominantly) equity market investors. We believe this is the first work to explore this topic from a macroeconomic, financial market dynamics and portfolio optimisation perspective.

In Section \ref{Economic_feature_similarity}, we conduct two sets of experiments. First, we study the evolution in the $L^1$ norms of inflation, economic growth and equity index return time series. This analysis confirms the recent history of each economic feature. Inflation is characterised by significant levels globally during the 1970s, followed by a steady decline and 20 years of relatively subdued levels throughout the period from 2000-2020. The most notable features in the GDP growth evolution are the precipitous drops corresponding to the GFC and COVID-19 market crash. The global equity index returns demonstrate a marked decline corresponding to the COVID-19 crash, also. In the second section, we aggregate all these time series in an effort to determine what the the economic driver is in the ultimate association between economic features. Our analysis indicates that country economic features group more predominantly with similar economic features, rather than other economic features belonging to the same country.

In Section \ref{Temporal_self_similarity_analysis}, we examine the temporal self-similarity exhibited by CPI inflation, GDP growth, and country equity index time series. Our analysis indicates variable temporal self-similarity among the features studied. The inflation distance matrix shows the highly anomalous nature of the elevated inflation levels throughout the 1970s, which preceded the relative homogeneity from the late 1990s onward. GDP growth and equity index returns both display broad temporal similarity, with the GFC and COVID-19 market crises displaying material difference in distance between all other candidate periods in time.

Next, we introduce a new algorithm to classify countries into one of four possible states corresponding to: i) ascending growth, ii) descending growth, iii) stagflation and iv) contraction. Having produced a time-varying economic state classification time series for each country, $\mathcal{S}_i(t)$, we compute an economic state integral $\Delta^{(i)}$, by averaging state classifications over time. Our new methods indicate that Japan's economy is most capable of exhibiting stagflationary/recessionary tendencies. We further investigate the collective similarity in these states among our collection of countries by computing a normalised inner product matrix, and then apply hierarchical clustering. This analysis confirms the existence of $k^{\mathcal{S}} = 2$ clusters, with Australia and Japan determined to be most dissimilar to the rest of the collection.

Finally, we implement a decade-by-decade portfolio optimisation to determine optimal country (and other) asset allocation for maximising portfolio Sharpe ratio. Our study demonstrates that optimal portfolio weights vary significantly from decade-to-decade, and this is largely driven by varying economic conditions experienced during different windows in time. One prominent example is the Japanese equity index, which averages an optimal capital allocation of 30.3\% during the first three decades, and declines to an average of .037\% in the final three decades. This reinforces the importance of continued financial portfolio monitoring and asset reallocation based on macroeconomic factors. The second primary finding is related to that of gold's importance in producing optimal risk-adjusted portfolio returns. Over the six decades studied, gold yields an average optimal allocation of $\sim$ 20.4\%. Furthermore, during the extreme inflation during the 1970s, and the two decades containing the GFC and COVID-19 market crash gold's optimal portfolio allocation was 40\%, 40\% and 35\% (noting that 40\% is the upper bound weight constraint throughout the optimisation experiments). These experiments demonstrate the importance gold may play in investor portfolios during a period of heightened economic uncertainty such as one we are now experiencing.

There are various opportunities for future work. First, one could study more economic variables over a longer period. Second, one could further develop the economic state classification algorithm and test the sensitivity for various thresholds. This algorithm is only an initial idea, and could be a basic building block for further statistical and mathematical modelling regarding economic regime classification. Finally, the portfolio optimisation section could be further developed to include further investor considerations such as portfolio leverage and shorting.

\bibliographystyle{_elsarticle-num-names}
\bibliography{__References.bib}
%\end{nolinenumbers}
\end{document}